\documentclass[review]{elsarticle}

\usepackage{lineno,hyperref}
\modulolinenumbers[2]
\usepackage{color, colortbl}
\usepackage{natbib}
\usepackage[margin=1in]{geometry}
\usepackage{bbding}
\usepackage{amsmath,amsfonts,amsthm,bm}
\usepackage{xcolor}
\usepackage{graphicx}

\definecolor{LightCyan}{rgb}{0.88,1,1}

\journal{Journal of \LaTeX\ Templates}

\begin{document}

\begin{frontmatter}





\title{AI challenges for predicting \\ the impact of mutations on protein stability}

\tnotetext[mytitlenote]{Fully documented templates are available in the elsarticle package on \href{http://www.ctan.org/tex-archive/macros/latex/contrib/elsarticle}{CTAN}.}

\author{Fabrizio Pucci}
\author{Martin Schwersensky}
\author{Marianne Rooman}
\address{Computational Biology and Bioinformatics, Université Libre de Bruxelles,\\ Brussels, Belgium\\
Interuniversity Institute of Bioinformatics in Brussels,\\ Brussels, Belgium}

\begin{abstract}
Stability is a key ingredient of protein fitness and its modification through targeted mutations has applications in various fields such as protein engineering, drug design and deleterious variant interpretation. Many studies have been devoted over the past decades to building new, more effective methods for predicting the impact of mutations on protein stability, based on the latest developments in artificial intelligence (AI). \textcolor{black}{We  discuss their features, algorithms, computational efficiency, and  accuracy estimated on an independent test set.  We focus on a critical analysis of their limitations, the recurrent  biases towards the training set, their generalizability and interpretability. We found that the accuracy of the predictors has stagnated at around 1 kcal/mol for over 15 years. We conclude by discussing the challenges that need to be addressed to reach improved performance.}
\end{abstract}


\end{frontmatter}

\section{Introduction}

The accurate prediction of mutational effects on protein stability is of utmost importance in many fields ranging from biotechnology to medicine. In rational protein engineering applications, for example, the targeted redesign of proteins makes it possible to optimize the biotechnological and biopharmaceutical processes in which they are involved \cite{korendovych2020novo,coluzza2017computational}.
Stability prediction also plays  a key role in interpreting  the impact of human genetic variants and may provide a better understanding of how these variants lead to disease conditions \cite{kopanos2019varsome,gunning2020assessing}. Note that stability is all the more important as it is the dominant factor in  protein fitness \cite{tokuriki2009stability}.

 For  these reasons,  many studies have been devoted over the last decade to the development of computational tools that aim to predict in a fast and reliable way the change in protein stability
 upon mutations \cite{dehouck2011popmusic,dehouck2009fast,savojardo2016inps,quan2016strum,capriotti2005mutant2,pires2014mcsm,pires2014duet,schymkowitz2005foldx,delgado2019foldx,kellogg2011role,cheng2006prediction,chen2020premps,li2020predicting,laimer2015maestro,cao2019deepddg,masso2014auto,huang2007iptree,witvliet2016elaspic,giollo2014neemo,chen2020istable,montanucci2019ddgun,benevenuta2021antisymmetric,li2021saafec}. These methods use information about protein sequence, structure and evolution,  which are combined through  a variety of machine learning methods ranging from simple linear regression   to  more complex models.  \textcolor{black}{For more information,  we refer to  excellent recent reviews \cite{sanavia2020limitations,marabotti2021predicting} and  comparative tests \cite{kepp2015towards,fang2020critical,iqbal2021assessing}}.
 
\textcolor{black}{ It has  to be noted} that, although  recent advances in the field of artificial intelligence
and more specifically in  deep learning have considerably improved feature selection and combination in multiple bioinformatics problems such as  three-dimensional (3D) protein structure prediction \cite{li2019deep,torrisi2020deep},  so far they are not often used in predicting  the effects of mutations  on protein stability. Indeed,   the majority of current predictors use shallow algorithms, probably because the amount of experimental training data is too limited to allow for deeper algorithms. 

 \textcolor{black}{In this  review, we concisely present the  protein stability prediction methods that are available and functional, and  test their performance on an independent set of  experimentally characterized point mutations, which are not part of any of the training sets. Our main goal here is to take a critical look at the predictors by investigating their algorithms,   limitations, and   biases. We also discuss} the main challenges the field will have to face in the  years to come in order to  strengthen the role of computational approaches in protein design and personalized medicine. 

\section{Brief overview \textcolor{black}{and benchmark} of the current computational models}

We collected \textcolor{black}{ 
existing computational methods  predicting  the change in protein thermodynamic  stability  upon point mutations,  defined by  the change in folding free energy $\Delta\Delta G$.
We restricted ourselves to predictors that are commonly used and   currently available through a working web server or downloadable code.}  These  methods,  listed  in Table \ref{AWA}, are almost all based on the 3D protein structure and use a  series of features such as the relative solvent accessible surface area (RSA) of the mutated residue, the change in folding free energy ($\Delta\Delta W$) estimated by various types of energy functions, the change in volume of the mutated residue  ($\Delta$Vol), and the change in residue hydrophobicity ($\Delta$Hyd). They also often use evolutionary information either extracted from  multiple sequence alignments of the query protein  or from substitution matrices such as BLOSUM62 \cite{henikoff1992blosum}.
 Several machine learning algorithms are used to combine  the different features. These are most often algorithms that have become classical   such as artificial neural networks, support vector machines or random forests. 
Only a few very recent predictors use novel deep learning approaches  \cite{li2020predicting,cao2019deepddg,benevenuta2021antisymmetric}. 
At the other extreme, a  predictor published this year  uses a  very simple model consisting of a linear combination of only three  features   \cite{caldararu2021three}. 

It is a difficult task to rigorously evaluate the accuracy of   predictors \cite{fang2020critical,iqbal2021assessing}. Indeed, performances depend on the training and test sets  as well as on the evaluation metric. \textcolor{black}{ Here, we have chosen  to  benchmark  the collected methods by  estimating their accuracy  in terms of} the root mean square error (RMSE) and the Pearson correlation coefficient ($r$) between experimental and predicted values for 830 mutations inserted in the 56-residue $\beta$1 extracellular domain of streptococcal protein G (PDB code 1PGA) \cite{nisthal2019protein}. \textcolor{black}{It has to be underlined  that this set of mutations is not included in the training sets of the  methods tested, and is thus a truly independent set.}

The RMSE of the predictors varies between 0.9 and 1.4 kcal/mol, and the correlation coefficients  between 0.3 and 0.7, as shown in Table \ref{AWA}. We observe low correlation between these two metrics: the method with the worst RMSE (1.42 kcal/mol) has the best $r$ (0.66). This follows from the fact that Pearson correlation coefficients are essentially driven by the points that are far from the mean, in contrast to RMSE which takes all points equally into account. 

\textcolor{black}{Note that these results must be interpreted with care. Indeed, both RMSE and $r$ values } depend on the distribution of  experimental $\Delta\Delta G$s  and more specifically, on its variance \cite{montanucci2019natural}. The ranking of the prediction methods and their scores thus crucially depend on the metric used and  on the test $\Delta\Delta G$ distribution. 

\textcolor{black}{In addition, we also tested  two other widely known stability predictors, FoldX \cite{delgado2019foldx} and Rosetta \cite{kellogg2011role}, which are physics-based rather than AI-based and employ full-atom representations rather than simplified descriptions of  protein structures. These two methods reach reasonable correlations with $r$ values of  0.36 and 0.44, respectively, slightly lower than AI-based methods ($\langle r\rangle=0.48$). In contrast, their RMSE values are above 3 kcal/mol, which is much worse than the average RMSE of 1.02 kcal/mol of AI-based methods. The lesser performance of these two methods has already been observed \cite{kepp2015towards} and could be due to the use of detailed atomic representation  which makes them sensitive to resolution defects.}

\section{ \textcolor{black}{Evolution of 
predictor performance over time}}

\begin{figure*}[h]
  \centering
\includegraphics[width=\textwidth]{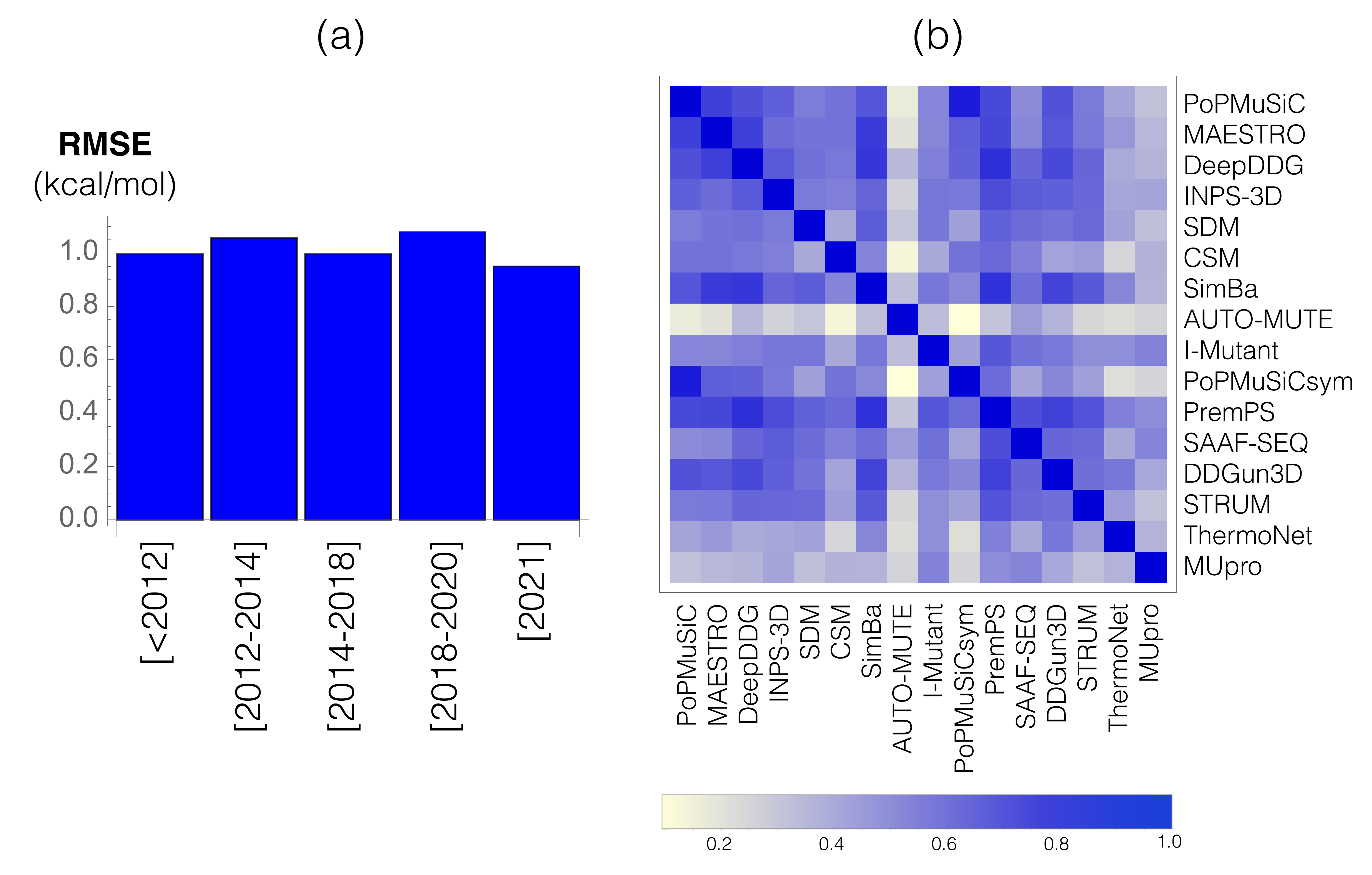}
\centering  
\caption{Evaluation of the $\Delta\Delta G$ prediction methods listed in Table \ref{AWA} on the basis of  the experimentally characterized mutations in the $\beta$1-extracellular domain of streptococcal protein G \cite{nisthal2019protein}. (a) Average RMSE of the predictors  as a function of their development date. (b) Correlation coefficients $r$ between the $\Delta\Delta G$s predicted by the different  methods.}
\label{time}
\end{figure*}

We have  analyzed the average performance of all the methods according to their year of development. We  clearly see in  Fig. \ref{time}.a that the average accuracy has not improved in the last 15 years, but basically remains constant, despite all efforts and the improved  performances claimed by the authors of   the newly published  methods. This is strikingly different from the situation in the field of protein structure prediction, for example, which has experienced an impressive improvement during the same period \cite{alquraishi2021machine}. 
Whether the accuracy limit 
on predicted $\Delta \Delta G$s is due to the relatively low number of mutations in the training set, to more fundamental reasons, or to uncontrolled biases in the predictors is currently a topic of debate  \cite{montanucci2019natural,caldararu2020systematic,sanavia2020limitations}. We discuss this issue  more extensively in the next sections. It must again be noted  that the RMSE threshold and the ranking of the methods performance can be somewhat different on other test mutations \cite{kepp2015towards,fang2020critical,iqbal2021assessing}. But the lower limit on RMSE is basically always around 1 kcal/mol.  

It is  instructive to 
look at  the correlations between the predictions of the different methods, shown in Fig. \ref{time}.b. They are all  reasonably good, with an average correlation coefficient of 0.5. This reflects that   the different methods use roughly the same information, but that there is   room for improvement and for further boosting the prediction accuracy by selecting informative features that have not yet been combined.

Another important characteristic of a prediction method is its speed. Indeed, as many current projects require investigating  protein stability properties at a large, proteome, scale \cite{schwersensky2020large},  the predictors have to be able to run fast enough to scan the proteome in a reasonable time.  All the methods tested are relatively fast with some  extremely fast such as  PoPMuSiC, SimBa, MAESTRO and AUTOMUTE (see Table \ref{AWA}). 

\begin{table}[h!]
\tabcolsep=0.11cm
\centering
\scalebox{0.8}{
\begin{tabular}{ccccccc}
\hline
	
\rowcolor{LightCyan} Method & 3D & Feature  &  RMSE  (kcal/mol)  & Run time& AI  & Ref. \\
\rowcolor{LightCyan} Year &  &  type &  $r$ & (min) &method &  \\\hline

\href{https://www.ics.uci.edu/~baldig/mutation.html}{MUpro} &  & Neighbors 
 & 1.17 & $<$ 1   &   Support vector   & \cite{capriotti2005mutant2} \\
(2006) & &  & $r$=0.26 & & regression  &  \\ \hline

\href{http://gpcr2.biocomp.unibo.it/cgi/predictors/I-Mutant3.0/I-Mutant3.0.cgi}{I-Mutant 3.0} &\Checkmark& Residue type,  RSA,
 & 0.92 & $\sim$ 400  &   Support vector   & \cite{capriotti2005mutant2} \\
(2007) & & Residue environment  & $r$=0.38 & & Regression  &  \\ \hline

\href{www.dezyme.com}{PoPMuSiC v2.1} &\Checkmark& Statistical potentials, & 0.95 & $<$ 1 &  Artificial neural  & \cite{dehouck2011popmusic} \\
(2011) & &$\Delta$Vol, RSA & $r$=0.56 & & network  &  \\ \hline

\href{http://marid.bioc.cam.ac.uk/sdm2}{SDM} &\Checkmark& RSA, Environment-specific  & 0.95 & $\sim$ 250 & Linear  & \cite{worth2011sdm} \\
(2011) && substitution frequencies& $r$=0.46 &&  combination&  \\ \hline

\href{http://biosig.unimelb.edu.au/mcsm}{mCSM} &\Checkmark& Graph-based
signatures,  &  1.10 & $\sim$ 250 &Regression via  & \cite{pires2014mcsm} \\
(2014) && Atomic distance patterns  & $r$=0.44& & Gaussian process   \\ \hline

\href{https://pbwww.services.came.sbg.ac.at/maestro/web}{MAESTRO} &\Checkmark& Statistical potentials, 
  & 0.91  &$<$ 1 &  Linear regression + & \cite{laimer2015maestro} \\
(2014) && PSize, ASA, SS, $\Delta$Hyd, $\Delta$IP   & $r$=0.58& & ANN + SVM   &\\ \hline

\href{http://binf.gmu.edu/automute/}{AUTOMUTE 2.0} &\Checkmark& 4-Body statistical potential
 & 1.16 &$\sim$ 1&   Random forest,  & \cite{masso2014auto} \\
(2014) && ASA, depth, SS, Vol   & $r$=0.30&& Tree regression &  \\ \hline


\href{https://inpsmd.biocomp.unibo.it/inpsSuite/default/index3D}{INPS-3D} &\Checkmark& Contact potential, RSA, EvolInfo,  & 0.96 & $\sim$4 &Support vector  & \cite{savojardo2016inps}\\ (2016)  &&  Bl62, $\Delta$Hyd, $\Delta$MW, MutI & $r$=0.52 & &  regression&\\ \hline

\href{https://zhanglab.ccmb.med.umich.edu/STRUM/}{STRUM} &\Checkmark& Energy functions, Homology modeling,  & 1.05 & $\sim$200 & Gradient boosting   & \cite{quan2016strum} \\ 
(2016) && $\Delta$Hyd, $\Delta$Vol, $\Delta$IP, $\Delta$MW,
EvolInfo&  $r$=0.49  && regression  &  \\ \hline

PoPMuSiC$^{\rm{sym}}$ &\Checkmark& Statistical potentials &  0.98 & $<$ 1 &  Artificial neural  & \cite{pucci2018quantification} \\
(2018) && $\Delta$Vol, RSA & $r$=0.54  & & network  &  \\ \hline


\href{https://github.com/biofold/ddgun}{DDGun3D} &\Checkmark& BL62, $\Delta$Hyd, RSA,  & 0.94 & $\sim$ 30 & Non-linear  &   \cite{montanucci2019ddgun} \\
(2019) &&  Statistical potentials & $r$=0.57  & &  regression &  \\ \hline

\href{http://protein.org.cn/ddg.html}{DeepDDG} &\Checkmark& ASA, SS, H-bonds, EvolInfo,  & 1.42 &$\sim$ 5 &  Shared residue pair   &   \cite{cao2019deepddg} \\
(2019) && Residue distances/orientations & $r$=0.66 & & deep neural network &  \\ \hline

\href{https://github.com/gersteinlab/ThermoNet}{ThermoNet} & \Checkmark &  Aromatic, Positive, Negative,  & 1.01  & $\sim$ 100 & 3D convolutional &   \cite{li2020predicting} \\
(2020) &&  Hyd, H-bond donor/acceptor& $r$=0.29 & 
 &  neural network &  \\ \hline

\href{https://lilab.jysw.suda.edu.cn/research/PremPS/}{PremPS} &\Checkmark&  EvolInfo, RSA, $\Delta$Hyd, Hyd, & 0.95 & $\sim$4 & Random  & \cite{chen2020premps} \\ 
(2020) &&  Aromatic, Charged, Leu& $r$=0.57 & &forest& \\ \hline

SimBa &\Checkmark& RSA, $\Delta$Vol & 0.99 & $<$ 1 & Linear  &   \cite{caldararu2021three} \\
(2021) && $\Delta$Hyd & $r$=0.53 &&   regression&  \\ \hline


\href{http://compbio.clemson.edu/SAAFEC-SEQ/}{SAAFEC-SEQ} & & EvolInfo, Neighbors, $\Delta$Vol,   & 0.91 & $\sim$ 30 & Gradient boosting &    \cite{li2021saafec}  \\
(2021) && $\Delta$Hyd, $\Delta$Flex, PSize, H-bond  & $r$=0.49 &&   decision tree & \\ \hline
\rowcolor{LightCyan}
& &$\langle$\textbf{RMSE}$\rangle$ =  & 1.02 $\pm$ 0.13 &&  &  \\ 
\rowcolor{LightCyan} && $\langle$\bf{$r$}$\rangle$ = & 0.48 $\pm$ 0.12 &&  &  \\
\rowcolor{LightCyan}&& $\bm \sigma${\bf(Exp}) =     & 0.98  &&  &  \\ 
 \hline
\end{tabular}
}
\caption{List of \textcolor{red}{ AI-based }$\Delta\Delta G$ predictors studied. The RMSE and linear correlation coefficient $r$ are computed for the experimentally characterized mutations in the $\beta$1-extracellular domain of streptococcal protein G \cite{nisthal2019protein}; $\sigma${(Exp}) is the standard deviation of the experimental $\Delta\Delta G$ distribution (in kcal/mol). Abbreviations used: ASA: solvent accessible surface area; RSA: relative ASA; Depth: surface, undersurface, or buried; PSize: Protein size; Vol: residue volume; $\Delta$Vol: Change in residue volume upon mutation; $\Delta$MW: Change in molecular weight; $\Delta$Flex: Change in flexibility; Hyd: Residue hydrophobicity; $\Delta$Hyd: Change in Hyd; SS: secondary structure; MutI:  mutability index of the native residue \cite{dayhoff1978mutability}; BL62: BLOSUM62 matrix \cite{henikoff1992blosum}; Neighbors: type of residues in the neighborhood along the sequence; EvolInfo: Evolutionary information from  protein families;
ANN: artificial neural network; SVM: supporting vector machine.} 
\label{AWA}
\end{table}









\section{Limitations and prediction biases}

  \begin{figure*}[h]
  \centering
\includegraphics[width=\textwidth]{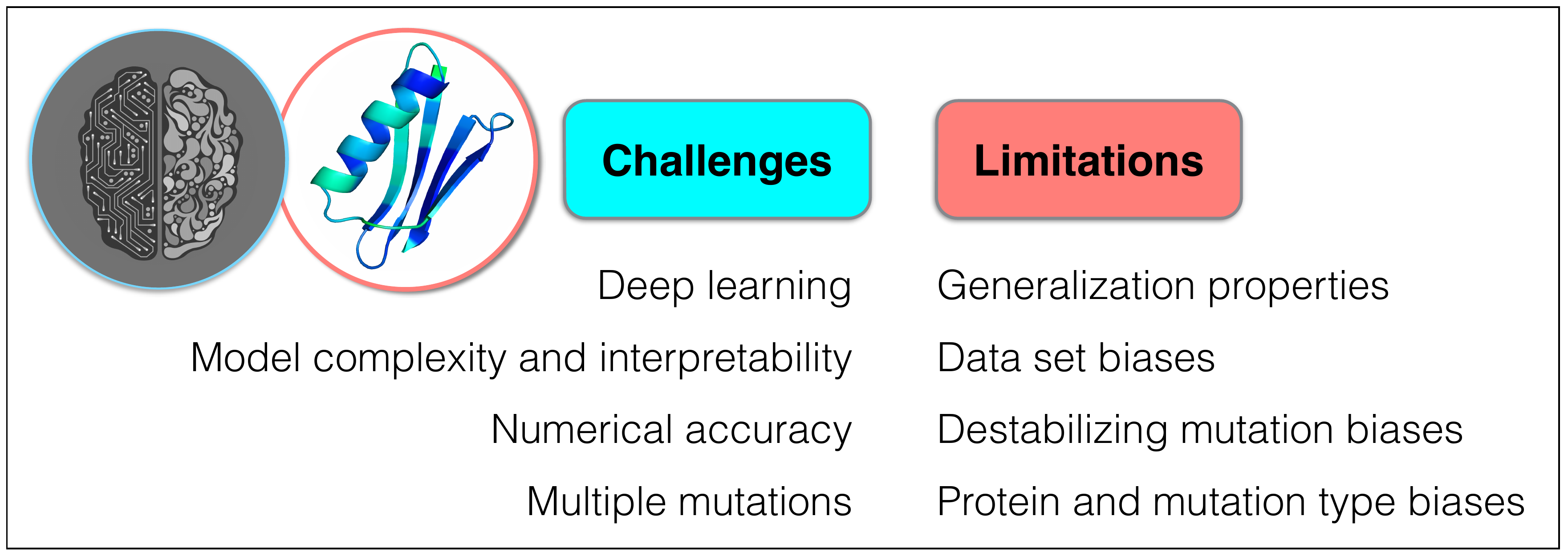}
\centering  
\caption{Schematic representation of the challenges and limitations that protein stability prediction methods have to address in the coming years.}
\label{MutaFrame}
\end{figure*}


The generalization property in  machine learning is the ability of the algorithm to correctly predict unseen data. The protein stability predictors, like all machine learning-based methods, tend however to be biased towards the data sets on which they are trained. \textcolor{black}{The majority of the methods analyzed here \cite{dehouck2009fast,pires2014mcsm,worth2011sdm,pucci2018quantification,savojardo2016inps,laimer2015maestro,caldararu2021three,chen2020premps,li2021saafec} were trained on the data set known as S2648 \cite{dehouck2009fast}. It contains 2,648 mutations with experimental $\Delta \Delta G$ values  collected from the literature and the ProTherm database \cite{gromiha2010protherm}, which were thoroughly checked and curated. Other predictors  use  subsets of S2648 or a slightly larger data set known as Q3421 \cite{quan2016strum}. }

Multiple hidden biases such as feature and hyperparameter selection biases that are difficult to control \textcolor{black}{can affect the generalization properties of the predictors trained on these data sets}. These problems are even more severe when complex algorithms are used or when the training sets are small and unbalanced. 
\textcolor{black}{In the following, we  quantitatively analyze a series of biases that often affect stability predictors  and are primarily caused by various imbalances in the training data sets, and discuss the strategies used to limit their impact.}  

\subsubsection*{\textbf{Cross-validation biases}}

Often, prediction performance is evaluated using a $k$-fold cross-validation procedure. 
This is  not always sufficient to estimate the accuracy of the methods and assessments on test sets are usually also provided, even though their sizes  are usually small. 
Going back to cross validation, there are different ways to perform the random split of the data set into $k$ folds:   at the level of the mutation, position,  protein  and even  protein cluster. Random splitting at mutation level introduces some distortions since the knowledge of the effect of a mutation at a given position makes the prediction of another substitution at the same position easier. Splitting at position level can also introduce some biases. To have more reliable estimations, cross validation at protein level has to be performed, or even at protein cluster level where all proteins that are similar to the target protein one wants to predict are removed from the training set. 

It should be noted that the extent to which the type of data set splitting affects  prediction performances is highly dependent on the prediction model. For example, the drop in performance of predictors that do not use complex machine learning like  PoPMuSiC and SimBa is almost negligible when passing from residue level to protein level  \cite{ancien2018SNP}.
In contrast, a substantial decrease in accuracy  is undergone by  STRUM,
with correlation coefficients and  RMSE between experimental and predicted    $\Delta \Delta G$s  that
pass from (0.77, 0.94 kcal/mol) for a 5-fold cross validation at mutation level to (0.64, 1.14 kcal/mol)  at position level and (0.54, 1.25 kcal/mol) at protein level \cite{quan2016strum}. A similar drop in performance of about 20-30\% when  strict cross validation procedures are employed has also been observed  in  \cite{chen2020premps}.

\subsubsection*{\textbf{Bias towards destabilizing mutations}}

At fixed environmental conditions,  the change in folding free energy upon mutation is antisymmetric by definition. More precisely, if  protein $B$ is a mutant of protein $A$, we have that $\Delta \Delta G (A\rightarrow B) = - \Delta \Delta G (B\rightarrow A)$. However, the majority of the stability predictors violate this relation, as shown by a series of studies \cite{pucci2015symmetry,pucci2018quantification,usmanova2018self,caldararu2020systematic}. This is mainly because training data sets are dominated by destabilizing mutations, which in turn results from the vast majority of mutations in a given protein being destabilizing.  For example, the ratio between the  numbers of destabilizing and stabilizing mutations in the data sets S2648  \cite{dehouck2009fast} and Q3421 \cite{quan2016strum}, \textcolor{black}{which are widely used as training sets,} are equal to  3.7 and 3.2, respectively, with a mean $\langle \Delta \Delta G\rangle$  of about 1 kcal/mol in both sets. 

Some of the recent prediction methods  got rid of this bias and satisfy the antisymmetry property by construction  \cite{chen2020premps,benevenuta2021antisymmetric,savojardo2016inps}. To check the extent to which  it is the case,  a balanced data set such as 
 $S^{\rm{sym}}$  \cite{pucci2018quantification} must be considered, which contains, for each mutation $A\rightarrow B$, the backward mutation $B\rightarrow A$ and thus an even number of stabilizing and destabilizing mutations. The deviation from  antisymmetry $\delta=\Delta \Delta G (A\rightarrow B) + \Delta \Delta G (B\rightarrow A)$ is an important measure for the evaluation of the lack of bias.

\subsubsection*{\textbf{Protein and mutation biases}}

Another type of bias arises from the fact that training data sets  do not provide a good sampling of the types of mutations and proteins, as recently discussed in \cite{caldararu2020systematic}. Often, mutation data sets are dominated by a few proteins which contain most of the entries and are therefore likely to bias the prediction towards them.  For example,  the 10 proteins from S2648 and Q3421 that contain the largest number of mutations represent 50\% and 40\%  of the  entries, respectively. 
The types of substitutions are also not well sampled:   among the 20$\times$19=380  possible amino acid substitutions, 78 and 38 are not sampled at all in S2648 and Q3421, respectively. The top 10 types are substitutions   into alanine, which account for 25\% of the entries in the  data sets. 

The way in which different methods are affected by this bias is extensively  evaluated in \cite{caldararu2020systematic} by introducing an unbiased test set with respect to mutation types. 
The majority of the  prediction methods are shown to be biased. They are able to correctly predict the effect of certain types of
mutations, while they completely miss others.

\section{Current and future challenges}

\subsection*{\textbf{Deep learning approaches}}

Deep learning algorithms such as such as convolutional neural networks have  provided spectacular improvements in a series of bioinformatics problems such as  protein structure prediction \cite{alquraishi2021machine}. Such methods are starting to be used in the prediction of the impact of mutations  on protein stability \cite{li2020predicting,cao2019deepddg,zhoudnpro,benevenuta2021antisymmetric}, but the majority of the current methods still use standard shallow machine learning approaches. This is due to the fact that  deep learning methods require large amounts of input data for training \cite{lecun2015deep}, while standard training data sets such as S2648 \cite{dehouck2011popmusic} or Q3421 \cite{quan2016strum}  only include a few thousand entries and are thus too small for these approaches. New mutation data  have recently been  collected \cite{nikam2021prothermdb,xavier2021thermomutdb,stourac2021fireprotdb}, which will certainly  increase the size of the training data sets after proper curation. However, these sets will probably remain too limited, with the consequence  that deep learning is  unlikely to outperform standard machine learning approaches without overfitting issues in the near future, even though unsupervised pre-training can help  prevent these issues to some extent \cite{lecun2015deep,benevenuta2021antisymmetric}. 

\subsection*{\textbf{Prediction model complexity and interpretability}}
The application of a wide variety of AI algorithms with different complexity to the prediction of protein stability is very informative.  These algorithms range from deep learning approaches such as  3D convolutional neural networks  \cite{li2020predicting} to extremely simple models such as  linear regression \cite{caldararu2021three}.
Complex algorithms can  capture  the intricate relationships between input features and training data better than simpler models, but they are in general more prone to overfitting. Moreover, most of them act as black boxes, which makes their results more difficult to interpret. 
Note that both over- and underfitting  are serious problems for generalization. Therefore, the development of a prediction model must be a trade-off between these two extremes. We would like to point out   that the best current methods are not always those that use the most complex AI techniques  (see Table \ref{AWA}).

The interpretability of the model at  biophysical and biochemical levels can be another characteristic to be considered in the model design. For example, it has been shown in \cite{caldararu2021three} that just three simple features, \emph{i.e.} the RSA of the mutated residues, and the change in residue volume and in hydrophobicity upon mutations, combined using a linear model, can achieve  performances similar to  state-of-the-art prediction methods that use up to hundred  features and complex machine learning. 
\textcolor{black}{Novel techniques for interpreting model predictions \cite{lundberg2017unified,AA,BB}, such as SHAP (SHapley Additive exPlanations) \cite{lundberg2017unified}, have recently been introduced in the AI field. Their application to protein stability predictors  helps to better identify the relative importance of features and to lead to more accurate prediction models  retaining interpretability properties.}

\subsection*{\textbf{Are we stuck with the limit of 1 kcal/mol RMSE ?}}

Surprisingly enough, all the methods developed over the past fifteen years have an accuracy evaluated by an RMSE slightly greater than 1 kcal/mol,  while most  validations on independent test sets are even worse with  RMSEs between 1.5 and 2.5 kcal/mol \cite{fang2020critical,iqbal2021assessing}. On the test protein we used  here, the situation is somewhat more favorable, with a lower value of 0.9  kcal/mol (Table \ref{AWA}); this is, however, related to the particularly low standard deviation of the  experimental $\Delta\Delta G$ distribution in this case (1 kcal/mol). 
The idea that 1 kcal/mol represents a hard limit for the prediction accuracy has already been suggested in \cite{caldararu2020systematic}. 

Several reasons can explain this limit. First, all the predictors are based  on a series of approximations, such as the use of  the wild type  structure but not the mutant structure. They  thus neglect the possible structural modifications caused by the mutations to the folded structure and, moreover,  also  overlook  perturbations to the unfolded state  \cite{caldararu2020systematic}. 
In addition,  entropy contributions to the folding free energy are largely overlooked, even though the methods based on statistical mean force potentials do not neglect them completely.
Another reason comes from the intrinsic errors on experimental $\Delta \Delta G$ values. In particular, both thermal and chemical measurements of $\Delta \Delta G$   generally involve approximations \cite{pucci2016high}. In addition, all the $\Delta \Delta G$ values in the data sets have not been determined under the same conditions, and the dependence of $\Delta \Delta G$ on, \emph{e.g.}, temperature or pH can be important. 

Whether the value of 1 kcal/mol is a true limit that cannot be circumvented, as suggested in \cite{montanucci2019natural,benevenuta2019upper} through a theoretical estimation of the experimental $\Delta \Delta G$   distribution and  noise,  is an open question. Our observation that the methods performance does not increase with time (Fig. \ref{time}.a) supports this view.   This question must be further investigated  to  understand if  and how  the current state-of-the-art predictors can be significantly improved. 

To address  these issues,  a systematic blinded experiment fully dedicated to the evaluation of protein stability changes upon mutations would  be of great benefit, in the same way that CASP (\href{https://predictioncenter.org/}{predictioncenter.org})  and CAPRI (\href{www.capri-docking.org}{capri-docking.org})  are for structure predictions.

\subsection*{\textbf{Metagenomic data}}

Metagenomic sequence data is a valuable source of sequence information that started to be used in protein structure prediction since the seminal paper of \cite{ovchinnikov2017protein}, and is now also extensively used in enzyme discovery \cite{D1NP00006C}. For example,  the majority of methods used such information as input in the last round of the CASP   experiment (CASP14) \cite{ovchinnikov2017protein}. 
Indeed, the enrichment of sequence data from metagenomic databases, even though they are often noisy, can improve protein sequence alignments and thus provide a more accurate assessment of how evolution shapes families of homologous proteins. 

Metagenomic sequence data is not yet used in the field of protein stability prediction, even not by the methods that have sequence conservation  among their features. This could be a way to  boost the prediction  accuracy.

\subsection*{\textbf{Multiple mutations versus single-point mutations}} 

Another challenge is to predict the effect of multiple mutations. It is of particular interest in protein design because multiple mutations can clearly lead to a higher degree of protein stabilization or destabilization  \cite{campeotto2017one,musil2017fireprot}. Yet, the vast majority of computational methods predict only the effect of single-site substitutions \cite{sanavia2020limitations}.  Point mutations can of course be combined to model multiple mutations, but this leads to neglecting any direct or indirect epistatic interactions between mutated residues \cite{schmiedel2019determining,rollins2019inferring}. 
The scarcity of experimental data on multiple mutations in a  variety of proteins as well as the degree of complexity compared to point mutations are the current limitations that prevent obtaining satisfactory prediction accuracy.



\bibliography{mybibfile}

\begin{thebibliography}{10}
\expandafter\ifx\csname url\endcsname\relax
  \def\url#1{\texttt{#1}}\fi
\expandafter\ifx\csname urlprefix\endcsname\relax\def\urlprefix{URL }\fi
\expandafter\ifx\csname href\endcsname\relax
  \def\href#1#2{#2} \def\path#1{#1}\fi

\bibitem{korendovych2020novo}
I.~V. Korendovych, W.~F. DeGrado, De novo protein design, a retrospective,
  Quarterly Reviews of Biophysics 53 (2020) e3.

\bibitem{coluzza2017computational}
I.~Coluzza, Computational protein design: a review, Journal of Physics:
  Condensed Matter 29~(14) (2017) 143001.

\bibitem{kopanos2019varsome}
C.~Kopanos, V.~Tsiolkas, A.~Kouris, C.~E. Chapple, M.~A. Aguilera, R.~Meyer,
  A.~Massouras, Varsome: the human genomic variant search engine,
  Bioinformatics 35~(11) (2019) 1978.

\bibitem{gunning2020assessing}
A.~C. Gunning, V.~Fryer, J.~Fasham, A.~H. Crosby, S.~Ellard, E.~L. Baple, C.~F.
  Wright, Assessing performance of pathogenicity predictors using clinically
  relevant variant datasets, Journal of Medical Genetics (2020)
  jmedgenet--2020--107003.

\bibitem{tokuriki2009stability}
N.~Tokuriki, D.~S. Tawfik, Stability effects of mutations and protein
  evolvability, Current opinion in structural biology 19~(5) (2009) 596--604.

\bibitem{dehouck2011popmusic}
Y.~Dehouck, J.~M. Kwasigroch, D.~Gilis, M.~Rooman, {PoPMuSiC} 2.1: a web server
  for the estimation of protein stability changes upon mutation and sequence
  optimality, BMC Bioinformatics 12~(1) (2011) 1--12.

\bibitem{dehouck2009fast}
Y.~Dehouck, A.~Grosfils, B.~Folch, D.~Gilis, P.~Bogaerts, M.~Rooman, Fast and
  accurate predictions of protein stability changes upon mutations using
  statistical potentials and neural networks: {PoPMuSiC-2.0}, Bioinformatics
  25~(19) (2009) 2537--2543.

\bibitem{savojardo2016inps}
C.~Savojardo, P.~Fariselli, P.~L. Martelli, R.~Casadio, {INPS-MD}: a web server
  to predict stability of protein variants from sequence and structure,
  Bioinformatics 32~(16) (2016) 2542--2544.

\bibitem{quan2016strum}
L.~Quan, Q.~Lv, Y.~Zhang, {STRUM}: structure-based prediction of protein
  stability changes upon single-point mutation, Bioinformatics 32~(19) (2016)
  2936--2946.

\bibitem{capriotti2005mutant2}
E.~Capriotti, P.~Fariselli, R.~Casadio, {I-Mutant2.0}: predicting stability
  changes upon mutation from the protein sequence or structure, Nucleic Acids
  Research 33~(suppl 2) (2005) W306--W310.

\bibitem{pires2014mcsm}
D.~E. Pires, D.~B. Ascher, T.~L. Blundell, {mCSM}: predicting the effects of
  mutations in proteins using graph-based signatures, Bioinformatics 30~(3)
  (2014) 335--342.

\bibitem{pires2014duet}
D.~E. Pires, D.~B. Ascher, T.~L. Blundell, {DUET}: a server for predicting
  effects of mutations on protein stability using an integrated computational
  approach, Nucleic Acids Research 42~(W1) (2014) W314--W319.

\bibitem{schymkowitz2005foldx}
J.~Schymkowitz, J.~Borg, F.~Stricher, R.~Nys, F.~Rousseau, L.~Serrano, The
  {FoldX} web server: an online force field, Nucleic Acids Research 33~(suppl
  2) (2005) W382--W388.

\bibitem{delgado2019foldx}
J.~Delgado, L.~G. Radusky, D.~Cianferoni, L.~Serrano, {FoldX 5.0}: working with
  {RNA}, small molecules and a new graphical interface, Bioinformatics 35~(20)
  (2019) 4168--4169.

\bibitem{kellogg2011role}
E.~H. Kellogg, A.~Leaver-Fay, D.~Baker, Role of conformational sampling in
  computing mutation-induced changes in protein structure and stability,
  Proteins: Structure, Function, and Bioinformatics 79~(3) (2011) 830--838.

\bibitem{cheng2006prediction}
J.~Cheng, A.~Randall, P.~Baldi, Prediction of protein stability changes for
  single-site mutations using support vector machines, Proteins: Structure,
  Function, and Bioinformatics 62~(4) (2006) 1125--1132.

\bibitem{chen2020premps}
Y.~Chen, H.~Lu, N.~Zhang, Z.~Zhu, S.~Wang, M.~Li, {PremPS}: Predicting the
  impact of missense mutations on protein stability, PLOS Computational Biology
  16~(12) (2020) e1008543.

\bibitem{li2020predicting}
B.~Li, Y.~T. Yang, J.~A. Capra, M.~B. Gerstein, Predicting changes in protein
  thermodynamic stability upon point mutation with deep {3D} convolutional
  neural networks, PLoS computational biology 16~(11) (2020) e1008291.

\bibitem{laimer2015maestro}
J.~Laimer, H.~Hofer, M.~Fritz, S.~Wegenkittl, P.~Lackner, {MAESTRO} -- multi
  agent stability prediction upon point mutations, BMC Bioinformatics 16~(1)
  (2015) 1--13.

\bibitem{cao2019deepddg}
H.~Cao, J.~Wang, L.~He, Y.~Qi, J.~Z. Zhang, {DeepDDG}: predicting the stability
  change of protein point mutations using neural networks, Journal of Chemical
  Information and Modeling 59~(4) (2019) 1508--1514.

\bibitem{masso2014auto}
M.~Masso, I.~I. Vaisman, {AUTO-MUTE} 2.0: a portable framework with enhanced
  capabilities for predicting protein functional consequences upon mutation,
  Advances in Bioinformatics 2014 (2014) 278385.

\bibitem{huang2007iptree}
L.-T. Huang, M.~M. Gromiha, S.-Y. Ho, {iPTREE-STAB}: interpretable decision
  tree based method for predicting protein stability changes upon mutations,
  Bioinformatics 23~(10) (2007) 1292--1293.

\bibitem{witvliet2016elaspic}
D.~K. Witvliet, A.~Strokach, A.~F. Giraldo-Forero, J.~Teyra, R.~Colak, P.~M.
  Kim, {ELASPIC} web-server: proteome-wide structure-based prediction of
  mutation effects on protein stability and binding affinity, Bioinformatics
  32~(10) (2016) 1589--1591.

\bibitem{giollo2014neemo}
M.~Giollo, A.~J. Martin, I.~Walsh, C.~Ferrari, S.~C. Tosatto, {NeEMO}: a method
  using residue interaction networks to improve prediction of protein stability
  upon mutation, BMC Genomics 15~(4) (2014) 1--11.

\bibitem{chen2020istable}
C.-W. Chen, M.-H. Lin, C.-C. Liao, H.-P. Chang, Y.-W. Chu, {iStable 2.0}:
  predicting protein thermal stability changes by integrating various
  characteristic modules, Computational and Structural Biotechnology Journal 18
  (2020) 622--630.

\bibitem{montanucci2019ddgun}
L.~Montanucci, E.~Capriotti, Y.~Frank, N.~Ben-Tal, P.~Fariselli, {DDGun}: an
  untrained method for the prediction of protein stability changes upon single
  and multiple point variations, BMC Bioinformatics 20~(14) (2019) 1--10\\
  $\bullet$ \emph{Method based on evolutionary information (DDGun) and
  additional structural information (DDGun3D) which predicts stability changes
  caused by point mutations but also by multiple mutations.}

\bibitem{benevenuta2021antisymmetric}
S.~Benevenuta, C.~Pancotti, P.~Fariselli, G.~Birolo, T.~Sanavia, An
  antisymmetric neural network to predict free energy changes in protein
  variants, Journal of Physics D: Applied Physics 54~(24) (2021) 245403.

\bibitem{li2021saafec}
G.~Li, S.~K. Panday, E.~Alexov, {SAAFEC-SEQ}: A sequence-based method for
  predicting the effect of single point mutations on protein thermodynamic
  stability, International Journal of Molecular Sciences 22~(2) (2021) 606.

\bibitem{sanavia2020limitations}
T.~Sanavia, G.~Birolo, L.~Montanucci, P.~Turina, E.~Capriotti, P.~Fariselli,
  Limitations and challenges in protein stability prediction upon genome
  variations: towards future applications in precision medicine, Computational
  and Structural Biotechnology Journal 18 (2020) 1968--1979.

\bibitem{marabotti2021predicting}
A.~Marabotti, B.~Scafuri, A.~Facchiano, Predicting the stability of mutant
  proteins by computational approaches: an overview, Briefings in
  Bioinformatics 22~(3) (2021) bbaa074.

\bibitem{kepp2015towards}
K.~P. Kepp, Towards a “golden standard” for computing globin stability:
  Stability and structure sensitivity of myoglobin mutants, Biochimica et
  Biophysica Acta (BBA)-Proteins and Proteomics 1854~(10) (2015) 1239--1248.

\bibitem{fang2020critical}
J.~Fang, A critical review of five machine learning-based algorithms for
  predicting protein stability changes upon mutation, Briefings in
  Bioinformatics 21~(4) (2020) 1285--1292.

\bibitem{iqbal2021assessing}
S.~Iqbal, F.~Li, T.~Akutsu, D.~B. Ascher, G.~I. Webb, J.~Song, Assessing the
  performance of computational predictors for estimating protein stability
  changes upon missense mutations, Briefings in Bioinformatics  bbab184.

\bibitem{li2019deep}
Y.~Li, C.~Huang, L.~Ding, Z.~Li, Y.~Pan, X.~Gao, Deep learning in
  bioinformatics: Introduction, application, and perspective in the big data
  era, Methods 166 (2019) 4--21.

\bibitem{torrisi2020deep}
M.~Torrisi, G.~Pollastri, Q.~Le, Deep learning methods in protein structure
  prediction, Computational and Structural Biotechnology Journal 18 (2020)
  1301--1310.

\bibitem{henikoff1992blosum}
S.~Henikoff, J.~G. Henikoff, Amino acid substitution matrices from protein
  blocks, Proceedings of the National Academy of Sciences 89~(22) (1992)
  10915--10919.

\bibitem{caldararu2021three}
O.~Caldararu, T.~L. Blundell, K.~P. Kepp, Three simple properties explain
  protein stability change upon mutation, Journal of Chemical Information and
  Modeling 61~(4) (2021) 1981--1988. \\$\bullet$ \emph{Simple model based on a
  linear combination of only three features (RSA, and change in hydrophobicity
  and volume upon mutation), which reaches stability change prediction scores
  similar to those of much more complex algorithms using hundreds of features}.

\bibitem{nisthal2019protein}
A.~Nisthal, C.~Y. Wang, M.~L. Ary, S.~L. Mayo, Protein stability engineering
  insights revealed by domain-wide comprehensive mutagenesis, Proceedings of
  the National Academy of Sciences 116~(33) (2019) 16367--16377.

\bibitem{montanucci2019natural}
L.~Montanucci, P.~L. Martelli, N.~Ben-Tal, P.~Fariselli, A natural upper bound
  to the accuracy of predicting protein stability changes upon mutations,
  Bioinformatics 35~(9) (2019) 1513--1517.

\bibitem{alquraishi2021machine}
M.~AlQuraishi, Machine learning in protein structure prediction, Current
  Opinion in Chemical Biology 65 (2021) 1--8\\ $\bullet$ \emph{Up--to--date
  review on protein structure prediction, which presents new ideas that could
  also be applied to boost the accuracy of predictors of protein stability
  changes upon mutation.}

\bibitem{caldararu2020systematic}
O.~Caldararu, R.~Mehra, T.~L. Blundell, K.~P. Kepp, Systematic investigation of
  the data set dependency of protein stability predictors, Journal of Chemical
  Information and Modeling 60~(10) (2020) 4772--4784 \\$\bullet$
  \emph{Extensive analysis of the impact of training data set properties on the
  accuracy of the prediction of protein stability changes upon mutations; the
  type of mutation, the extent of stabilization, the type of structure and the
  solvent exposure are carefully analyzed as possible sources of bias }.

\bibitem{schwersensky2020large}
M.~Schwersensky, M.~Rooman, F.~Pucci, Large-scale in silico mutagenesis
  experiments reveal optimization of genetic code and codon usage for protein
  mutational robustness, BMC biology 18~(1) (2020) 1--17.

\bibitem{worth2011sdm}
C.~L. Worth, R.~Preissner, T.~L. Blundell, {SDM} -- a server for predicting
  effects of mutations on protein stability and malfunction, Nucleic Acids
  Research 39~(suppl\_2) (2011) W215--W222.

\bibitem{pucci2018quantification}
F.~Pucci, K.~V. Bernaerts, J.~M. Kwasigroch, M.~Rooman, Quantification of
  biases in predictions of protein stability changes upon mutations,
  Bioinformatics 34~(21) (2018) 3659--3665\\$\bullet$ \emph{In--depth study of
  the bias towards destabilizing mutations, quantified through the introduction
  of a new balanced mutation data set; almost all stability predictors are
  shown to suffer from this bias}.

\bibitem{dayhoff1978mutability}
M.~Dayhoff, R.~Schwartz, B.~Orcutt, Atlas of Protein Sequence and Structure,
  National Biomedical Research Foundation, Washington DC, 1978, Ch. A model of
  evolutionary change in proteins, pp. 345--352.

\bibitem{gromiha2010protherm}
M.~M. Gromiha, A.~Sarai, Thermodynamic database for proteins: features and
  applications, Methods in Molecular Biology 609 (2010) 97--112.

\bibitem{ancien2018SNP}
F.~Ancien, F.~Pucci, M.~Godfroid, M.~Rooman, Prediction and interpretation of
  deleterious coding variants in terms of protein structural stability,
  Scientific reports 8 (2018) 4480.

\bibitem{pucci2015symmetry}
F.~Pucci, K.~Bernaerts, F.~Teheux, D.~Gilis, M.~Rooman, Symmetry principles in
  optimization problems: an application to protein stability prediction,
  IFAC-PapersOnLine 48~(1) (2015) 458--463.

\bibitem{usmanova2018self}
D.~R. Usmanova, N.~S. Bogatyreva, J.~Ari{\~n}o~Bernad, A.~A. Eremina, A.~A.
  Gorshkova, G.~M. Kanevskiy, L.~R. Lonishin, A.~V. Meister, A.~G. Yakupova,
  F.~A. Kondrashov, et~al., Self-consistency test reveals systematic bias in
  programs for prediction change of stability upon mutation, Bioinformatics
  34~(21) (2018) 3653--3658 \\$\bullet$ \emph{Analysis of the prediction bias
  towards destabilizing mutations of a series of predictors of protein
  stability changes upon mutations, which are all shown to be biased.}

\bibitem{zhoudnpro}
X.~Zhou, J.~Cheng, {DNpro}: A deep learning network approach to predicting
  protein stability changes induced by single-site mutations, Journal of
  Bioengineering and Life Sciences 10~(5) (2016) 1--7.

\bibitem{lecun2015deep}
Y.~LeCun, Y.~Bengio, G.~Hinton, Deep learning, Nature 521~(7553) (2015)
  436--444.

\bibitem{nikam2021prothermdb}
R.~Nikam, A.~Kulandaisamy, K.~Harini, D.~Sharma, M.~M. Gromiha, {ProThermDB}:
  thermodynamic database for proteins and mutants revisited after 15 years,
  Nucleic Acids Research 49~(D1) (2021) D420--D424 \\$\bullet$ \emph{New
  manually curated database reporting experimental data on the impact of
  mutations on the thermal and thermodynamic stability of proteins}.

\bibitem{xavier2021thermomutdb}
J.~S. Xavier, T.-B. Nguyen, M.~Karmarkar, S.~Portelli, P.~M. Rezende, J.~P.
  Velloso, D.~B. Ascher, D.~E. Pires, {ThermoMutDB}: a thermodynamic database
  for missense mutations, Nucleic Acids Research 49~(D1) (2021) D475--D479
  \\$\bullet$ \emph{New manually curated database reporting experimental data
  on the impact of mutations on the thermal and thermodynamic stability of
  proteins}.

\bibitem{stourac2021fireprotdb}
J.~Stourac, J.~Dubrava, M.~Musil, J.~Horackova, J.~Damborsky, S.~Mazurenko,
  D.~Bednar, {FireProtDB}: database of manually curated protein stability data,
  Nucleic acids research 49~(D1) (2021) D319--D324 \\$\bullet$ \emph{New
  manually curated database reporting experimental data on the impact of
  mutations on the thermal and thermodynamic stability of proteins}.

\bibitem{lundberg2017unified}
S.~M. Lundberg, S.-I. Lee, A unified approach to interpreting model
  predictions, in: Proceedings of the 31st international conference on neural
  information processing systems, 2017, pp. 4768--4777.

\bibitem{AA}
A.~Shrikumar, P.~Greenside, A.~Kundaje, Learning important features through
  propagating activation differences, in: D.~Precup, Y.~W. Teh (Eds.),
  Proceedings of the 34th International Conference on Machine Learning, Vol.~70
  of Proceedings of Machine Learning Research, PMLR, 2017, pp. 3145--3153.

\bibitem{BB}
E.~{\v{S}}trumbelj, I.~Kononenko, Explaining prediction models and individual
  predictions with feature contributions, Knowledge and information systems
  41~(3) (2014) 647--665.

\bibitem{pucci2016high}
F.~Pucci, R.~Bourgeas, M.~Rooman, High-quality thermodynamic data on the
  stability changes of proteins upon single-site mutations, Journal of Physical
  and Chemical Reference Data 45~(2) (2016) 023104.

\bibitem{benevenuta2019upper}
S.~Benevenuta, P.~Fariselli, On the upper bounds of the real-valued
  predictions, Bioinformatics and Biology Insights 13 (2019) 1177932219871263.

\bibitem{ovchinnikov2017protein}
S.~Ovchinnikov, H.~Park, N.~Varghese, P.-S. Huang, G.~A. Pavlopoulos, D.~E.
  Kim, H.~Kamisetty, N.~C. Kyrpides, D.~Baker, Protein structure determination
  using metagenome sequence data, Science 355~(6322) (2017) 294--298.

\bibitem{D1NP00006C}
S.~L. Robinson, J.~Piel, S.~Sunagawa, A roadmap for metagenomic enzyme
  discovery, Natural Product Reports (2021) --\href
  {http://dx.doi.org/10.1039/D1NP00006C} {\path{doi:10.1039/D1NP00006C}}.

\bibitem{campeotto2017one}
I.~Campeotto, A.~Goldenzweig, J.~Davey, L.~Barfod, J.~M. Marshall, S.~E. Silk,
  K.~E. Wright, S.~J. Draper, M.~K. Higgins, S.~J. Fleishman, One-step design
  of a stable variant of the malaria invasion protein rh5 for use as a vaccine
  immunogen, Proceedings of the National Academy of Sciences 114~(5) (2017)
  998--1002.

\bibitem{musil2017fireprot}
M.~Musil, J.~Stourac, J.~Bendl, J.~Brezovsky, Z.~Prokop, J.~Zendulka,
  T.~Martinek, D.~Bednar, J.~Damborsky, {FireProt}: web server for automated
  design of thermostable proteins, Nucleic Acids Research 45~(W1) (2017)
  W393--W399.

\bibitem{schmiedel2019determining}
J.~M. Schmiedel, B.~Lehner, Determining protein structures using deep
  mutagenesis, Nature Genetics 51~(7) (2019) 1177--1186.

\bibitem{rollins2019inferring}
N.~J. Rollins, K.~P. Brock, F.~J. Poelwijk, M.~A. Stiffler, N.~P. Gauthier,
  C.~Sander, D.~S. Marks, Inferring protein {3D} structure from deep mutation
  scans, Nature Genetics 51~(7) (2019) 1170--1176.

\end{thebibliography}

\end{document}